\documentclass[11pt]{article}
\usepackage[colorlinks]{hyperref}
\hypersetup{
    urlcolor=blue
}
\usepackage{amssymb,amsbsy}
\usepackage{amsfonts,amsbsy,mathrsfs}
\usepackage{amssymb}
\usepackage{amsfonts}
\usepackage{amsmath}%

\newcommand{\be}{\begin{equation}}
\newcommand{\ee}{\end{equation}}
\newcommand{\ba}{\begin{array}}
\newcommand{\ea}{\end{array}}

\newcommand{\bu}{\boldsymbol{u}}

\newtheorem{thm}{Theorem}

\newtheorem{prop}{Proposition}

\newtheorem{example}{Example}

\date{}
\begin{document}

\title{\protect\vspace*{-15mm}
{\bf Multidimensional integrable systems 
from contact geometry}\protect\vspace*{-0.2cm}
}

\author{{\large \sc Artur Sergyeyev}\\Mathematical Institute, Silesian University in Opava, \\ Na Rybn\'\i{}\v{c}ku 1, 74601 Opava, Czech Republic\\ 
ORCID: \href{https://orcid.org/0000-0001-8394-6165}{0000-0001-8394-6165}\thanks{This version of the article is based on the accepted manuscript and reflects, to the extent deemed reasonable by the author, the post-acceptance corrections, and, while it differs in formatting from the published Version of Record, some flaws of the latter are fixed in the present version, which also offers some additional enhancements like fully linked bibliography. The Version of Record is available online at
\url{http://dx.doi.org/10.1007/s40590-024-00703-7} and is published under the Creative Commons license, see \url{http://creativecommons.org/licenses/by/4.0/} for details. The reference to the published version is: A. Sergyeyev, Multidimensional integrable systems from contact geometry, {\em Bol. Soc. Mat. Mex.} {\bf 31}  (2025), art. 26. 
}
}

\maketitle\thispagestyle{empty}

\begin{abstract}\protect\vspace*{-1.2cm}
Upon having presented a bird's eye view of history of integrable systems, we give a brief review of certain recent advances in the longstanding problem of search for partial differential systems in four independent variables, often referred to as (3+1)-dimensional or 4D systems, that are integrable in the sense of soliton theory. Namely, we review a recent construction for a large new class of (3+1)-dimensional integrable systems with Lax pairs involving contact vector fields. This class contains inter alia two infinite families of such systems, thus establishing that there is significantly more integrable (3+1)-dimensional systems than it was believed for a long time. 

In fact, the construction under study yields (3+1)-dimensional integrable generalizations of many well-known dispersionless integrable (2+1)-dimensional systems like the dispersionless KP equation, as well as a first example of a (3+1)-dimensional integrable system with an algebraic, rather than rational, nonisospectral Lax pair.

To demonstrate the versatility of the construction in question, we employ it here to produce novel integrable (3+1)-dimensional generalizations for the following (2+1)-dimensional integrable systems: dispersionless BKP, dispersionless asymmetric Nizhnik--Veselov--Novikov, dispersionless Gardner, and dispersionless modified KP equations, and the generalized Benney system.\looseness=-1 

{\bf Keywords:} (3+1)-dimensional integrable systems; 
Lax pairs; contact geometry

{\bf MSC 2020:} 37K10; 17B80
\end{abstract}
\protect\vspace*{-7mm}
\section{Introduction}

The theory of integrable systems has its roots in trying to answer a simple and natural question: when
an ordinary differential equation, or a system of such equations, can be integrated by quadratures? An important milestone here is the Liouville theorem \cite{L} in classical mechanics giving a sufficient condition for this to occur for an important class of integrable Hamiltonian systems that are, inter alia, of considerable significance for applications, see e.g.\ \cite{arn}; 
such systems were intensively studied by the researchers from Mathematical Institute of UNAM, see e.g.\ recent works \cite{At1, At2} and references therein.
\looseness=-1

The next great breakthrough in the field of integrability has occurred about half a century ago with the discovery of the so-called inverse scattering transform (IST) in the seminal work \cite{GGKM}. In 
\cite{GGKM} it was shown that solving the Cauchy problem for the {\em nonlinear} Korteweg--de Vries (KdV) equation, a remarkable evolutionary partial differential equation in one dependent and two independent variables satisfied, inter alia, by a certain generating function for intersection numbers of complex curves arising in the Witten conjecture and its proof by Kontsevich, see e.g.\ \cite{Tau}, can, under certain conditions, be reduced to a sequence of {\em linear} problems, and the procedure in question became known as IST. This is possible because the KdV equation can be written as a compatibility condition for an overdetermined system of linear equations; such overdetermined systems in the context of integrable systems are called the Lax pairs or the Lax/Lax-type representations,  
see e.g.\ \cite{Dun, ir, lax, lwy, w} for details. An important consequence of the above
is the construction of infinitely many explicit exact solutions of the KdV equation -- the multisoliton solutions, see e.g.\ \cite{ir, w} for more details on the latter.\looseness=-1

Moreover, existence of a Lax pair for the KdV paved the way to construction \cite{DMN} of other important classes of explicit exact solutions for this equation, namely, the quasiperiodic finite-gap solutions that are inextricably related to algebraic geometry. Note that making use of the counterparts of these solutions for the KP equation, a natural integrable generalization of the KdV equation to the case of three independent variables, made it possible to prove \cite{Shi} the longstanding Schottky conjecture in algebraic geometry.\nopagebreak[4]

Yet another notable consequence of the presence of the Lax pair is existence of infinitely many nontrivial independent local conservation laws for the KdV equation, see e.g.\ \cite{ir, lwy, Magri} and references therein, which shows, inter alia, that the associated dynamics is highly regular rather than chaotic.

It was quickly realized that the KdV equation is by no means an isolated example -- there is plenty of partial differential systems that admit `good' Lax pairs from which infinite hierarchies of conservation laws can be extracted and that are, at least in principle, amenable to the IST.
In what follows we shall call {\em integrable} 
the partial differential systems with `good' Lax pairs in the above sense, unlike some authors who use the term integrable 
also for the systems that can be linearized by an appropriate change of variables like e.g.\ the Burgers equation. 

Note that soliton and multisoliton solutions for KdV and many other integrable systems, as well as other types of exact solutions constructed using the Lax pairs, like 
e.g.\ 
the multi-instanton solutions for the (anti)self-dual Yang--Mills equations obtained using the 
famous ADHM construction \cite{ADHM}, have found significant applications both in physics and in pure mathematics, see for example Donaldson's revolutionary works on geometry of four-dimensional manifolds using instantons, see e.g.\ \cite{At}.\looseness=-1

Integrable systems are well known to have a number of remarkable structures attached to them.
These include Lax pairs, symmetries, conservation laws, Hamiltonian 
structures and more, see e.g.\ \cite{fc, Dun, ir, kvv, lwy, Magri, olv, s, t}.
Exploring these structures and their properties can provide one with quite a bit of insight into the behavior of the systems under study and their properties, be these systems integrable or not. For one, the presence of large number of symmetries and/or conservation laws indicates that the system under study has a highly constrained, and hence likely quite regular, dynamics, see e.g.\ the discussion in \cite{arn, ir}.\looseness=-1

Among integrable partial differential systems, those in four independent variables (often referred to as (3+1)-dimensional ones even if all four variables are on equal footing) are of particular interest, as four is the dimension of our spacetime according to general relativity, so gaining a deeper understanding of such systems could be quite significant for possible applications.\looseness=-1

Integrable (3+1)-dimensional systems were long believed to be quite scarce, with the 
known examples being mostly related in some way to the concept of (anti)self-duality, as is the case e.g.\ for the two most important ones, (anti)self-dual Yang--Mills equations and 
the (anti)self-dual vacuum Einstein equations; 
an effective construction for integrable (3+1)-dimensional systems appeared rather elusive. 

It was, however, recently shown that this is not the case and there is  \cite{aslmp} an effective construction that produces a large new class of integrable (3+1)-dimensional systems. We review this construction in detail, along with a number of new examples, in the next section.


The integrable (3+1)-dimensional systems resulting from the construction in question, as well as the overwhelming majority of previously known ones, are 
dispersionless in the following sense.

A partial differential system is said to be {\em dispersionless}, or of {\em hydrodynamic type},  see e.g.\ \cite{kod, aslmp, w, z} and references therein, if it can be written as a first-order homogeneous quasilinear system, that is,
\begin{equation}\label{dsgen}
A_1(\bu) \bu_{x^1}+A_2(\bu)\bu_{x^2}+\cdots+A_{d}(\bu)\bu_{x^{d}}=0;
\end{equation}
here $d$, $M$ and $N$ are natural numbers, $A_i$~are~$M\times N$~matrices, $M\geqslant N$,~$\bu\equiv(u^1,\dots,u^N)^\mathrm{T}$ is the vector of unknown functions and
$\vec{x}=(x^1,\dots,x^{d})^\mathrm{T}$ of independent variables, so $\bu=\bu(\vec{x})$; the superscript $\mathrm{T}$ here and below indicates the transposed matrix. In what follows all functions will be ta
citly assumed sufficiently smooth for all computations to make sense.

Dispersionless systems have many applications in 
fluid dynamics, which largely is the motivation behind the term hydrodynamic-type systems, and in nonlinear optics and general relativity, see for example \cite{bsz, h, aslmp, tt, w, z} and references therein.

Below we shall deal with dispersionless systems in four independent variables (i.e., (3+1)-dimensional ones); 
the independent variables will be denoted $x,y,z,t$ (thus from now on $\vec x=(x,y,z,t)^\mathrm{T}$), so the systems under study 
read\looseness=-1
\begin{equation}\label{sys-gen}
A_1(\bu)\bu_x+A_2(\bu)\bu_y+A_3(\bu)\bu_z+A_4(\bu)\bu_t=0. 
\end{equation}

In keeping with the above remark that the overwhelming majority of integrable (3+1)-dimensional systems known to date can be written in dispersionless form, 
let us show that this is the case for the anti-self-dual Yang--Mills equations with a matrix Lie group as a gauge group on $\mathbb{R}^{4}$ with metric with neutral signature, the (3+1)-dimensional integrable system of utmost importance for physics.\looseness=-1

Indeed, it is known, cf.\ e.g.\ \cite{asoro} and references therein, that upon a proper choice of gauge and of local coordinates on $\mathbb{R}^{4}$ these equations boil down to a single equation for the so-called Yang matrix $J$\looseness=-1
\[
\partial_{z} (J^{-1}\partial_x J)+ \partial_{t} (J^{-1}\partial_y J)=0,
\]
that upon introducing another matrix $K$ can be rewritten in 
dispersionless form as
\begin{equation}\label{jk}
\partial_x J= J \partial_{t} K,\quad \partial_y J= -J \partial_{z} K.
\end{equation}  

\section{Integrable (3+1)-dimensional systems with contact Lax pairs}

For an $h=h(p,\bu)$ define 
an operator $X_h$ as 
\begin{equation}\label{cvf}
X_h=h_p\partial_x+(ph_z-h_x)\partial_p+(h-p h_p)\partial_z
\end{equation}
which formally looks exactly like the contact vector field with a contact Hamiltonian $h$ on a contact 3-manifold with local coordinates $x,z,p$ and contact one-form $dz+pdx$, see \cite{aslmp} 
for details; cf.\ also e.g.\ \cite{arn, bra} and references therein for general background on contact geometry.

A 
linear
system of the form \cite{aslmp}
\begin{equation}\label{linearlax-0}
\chi_y=X_{f}(\chi),\quad \chi_t=X_{g}(\chi)
\end{equation}
for $\chi=\chi(x,y,z,t,p)$ will be hereinafter referred to as a 
{\em (linear) contact Lax pair}. 
Here $p$~is~the so-called variable spectral parameter (recall that  $\bu=\bu(x,y,z,t)$, so 
$\bu_p\equiv 0$), $f=f(p,\bu)$, $g=g(p,\bu)$ are the {\em  Lax functions} and $L=\partial_y-X_f$ and $M=\partial_t-X_g$ are the {\it Lax operators}.

The Lax pairs (\ref{linearlax-0}) provide a new and natural {(3+1)-dimensional} generalization of the well-known, see e.g.\ \cite{aslmp, z} and references therein, {(2+1)-dimensional} Lax pairs
\begin{equation}\label{3dllax}
\chi_y=\mathcal{X}_f(\chi), \quad\chi_t=\mathcal{X}_g(\chi),
\end{equation}
where $\mathcal{X}_h=h_p \partial_x-h_x\partial_p$,
since if
$\bu_z=0$ and $\chi_z=0$ then (\ref{linearlax-0}) boils down to (\ref{3dllax}).
The class of integrable (2+1)-dimensional systems with Lax pairs (\ref{3dllax}) is known to be quite broad, see e.g.\ \cite{fkh, aslmp, z} and references therein,  so it is natural to ask whether the same holds true for the class of integrable (3+1)-dimensional systems with linear contact Lax pairs (\ref{linearlax-0}).\looseness=-1

The following result, summarizing the key findings of \cite{aslmp}, shows that this is indeed the case and there 
are infinitely many pairs $(f,g)$ such that
the systems for $\bu$ admitting 
Lax pairs of the form (\ref{linearlax-0}) are genuinely (3+1)-dimensional integrable nonlinear systems 
transformable into Cauchy--Kowalevski form.

\begin{thm}
\label{conth}
Contact Lax~pairs (\ref{linearlax-0})
yield 
integrable {(3+1)-dimensional}
systems
that can be brought
into Cauchy--Kowalevski form {\em inter alia} for the following two infinite sets of pairs of Lax functions $f$ and $g$:\looseness=-1

I.
$f=p^{m+1}+\displaystyle\sum\limits_{i=0}^{m} u_i p^i,\quad 
g=p^{n+1}+{\displaystyle\frac{n}{m}} u_{m} p^{n}+\displaystyle\sum\limits_{j=0}^{n-1} v_j p^j$ 
with $\bu=(u_0,\dots,u_m,v_0, \dots,v_{n-1})^\mathrm{T}$,

II. $f=\sum\limits_{i=1}^{m} \displaystyle\frac{a_i}{(p-u_i)},\quad
g=\sum\limits_{j=1}^{n}\displaystyle\frac{b_j}{(p-v_j)}$ 
with $\bu=(a_1,\dots,a_{m}, u_1,\dots,u_{m},b_1,\dots,b_{n}, v_1,\dots,v_{n})^\mathrm{T},$\looseness=-1
\bigskip

\noindent where 
$m$ and $n$ are arbitrary natural numbers.
\end{thm}

Before proceeding to the proof of the above theorem, 
consider for a moment the {\em Lax~equation}
\begin{equation}\label{laxeq}
[\partial_y-X_f, \partial_t-X_g]=0,
\end{equation}
expressing the compatibility condition for (\ref{linearlax-0}).
This compatibility condition can be expressed in a more concise form that simplifies many computations:
\begin{prop}[\cite{aslmp}]
The Lax equation (\ref{laxeq}) holds iff so does
a~zero-curvature-type~equation
\begin{equation}\label{zcr}
f_t- g_y+\{ f, g\}=0, 
\end{equation}
where the bracket $\{,\}$ is the contact bracket in dimension three, namely, 
\begin{equation}\label{c-b}\{f,g\}=f_{p} g_x-g_{p} f_x
 -p\left(f_{p} g_z-g_{p} f_z\right)+ f g_z-g f_z
\end{equation} 
\end{prop}
\noindent{\em Sketch of proof of Theorem~\ref{conth}.} Straightforward but tedious computations \cite{aslmp} show that for Case I the compatibility condition (\ref{zcr}) for (\ref{linearlax-0}), upon equating to zero the coefficients at all powers of $p$, yields the system\looseness=-1 
\begin{equation}\label{cli}
\hspace*{-5mm}
\begin{array}{l}
\displaystyle \left(u_k\right)_t-\left(v_k\right)_y+n\left(u_{k-n-1}\right)_z-m\left(v_{k-m-1}\right)_z
+(m+1)\left(v_{k-m}\right)_x-(n+1)\left(u_{k-n}\right)_x
\\[3mm]
+\displaystyle\sum\limits_{i=0}^{m}\biggl\{(k-i-1)v_{k-i} \left(u_i\right)_z-(i-1) u_i \left(v_{k-i}\right)_z
-(k+1-i)v_{k+1-i} \left(u_i\right)_x+i u_i \left(v_{k+1-i}\right)_x\biggr\}=0,\\[7mm] 
k=0,\dots,n+m, 
\end{array}
\end{equation}
where 
to simplify writing we have adopted 
a notational convention that 
$u_{m+1}\stackrel{\mathrm{def}}{=}1$, $v_{n+1}\stackrel{\mathrm{def}}{=}1$,  $u_i\stackrel{\mathrm{def}}{=}0$ for $i>m+1$ and $i<0$, $v_j\stackrel{\mathrm{def}}{=}0$ for $j>n+1$ and $j<0$, and finally $v_n\stackrel{\mathrm{def}}{=}(n/m) u_m$.

It can be shown \cite{aslmp} that (\ref{cli}) is an evolution system in disguise, and hence indeed can be 
brought into the Cauchy--Kowalevski form: namely, (\ref{cli})  
can be solved w.r.t.\  
the derivatives $(u_i)_z$ and $(v_j)_z$ for all $i=0,\dots,m$ and $j=0,\dots,n-1$.\looseness=-1

Likewise, for 
Case II 
we have \cite{aslmp} that the compatibility condition (\ref{zcr}) for (\ref{linearlax-0}), upon being brought to the common denominator and subsequent equating to zero the coefficients at all powers of $p$ in the numerator, yields the system 
\begin{equation}\label{claii}
\begin{array}{rl}
(u_i)_t&+\displaystyle\sum\limits_{j=1}^n\left\lbrace \left(\displaystyle\frac{b_j}{v_j-u_i}\right)_x-\left(\displaystyle\frac{b_j u_i}{v_j-u_i}\right)_z+\displaystyle\frac{2 b_j (u_i)_z}{v_j-u_i}\right\rbrace=0,\quad i=1,\dots,m,\\[6mm]
(v_j)_y&+\displaystyle\sum\limits_{i=1}^m\left\lbrace -\left(\displaystyle\frac{a_i}{v_j-u_i}\right)_x+\left(\displaystyle\frac{a_i v_j}{v_j-u_i}\right)_z-\displaystyle\frac{2 a_i (v_j)_z}{v_j-u_i}\right\rbrace=0,\quad j=1,\dots,n,\displaybreak[4]\\[7mm]
(a_i)_t&+\displaystyle\sum\limits_{j=1}^n\left\lbrace \left(\displaystyle\frac{a_i b_j}{(v_j-u_i)^2}\right)_x+\left(\displaystyle\frac{a_i b_j (v_j-2 u_i)}{(v_j-u_i)^2}\right)_z
\right.\\[6mm]
&\left.
-\displaystyle\frac{3 a_i (b_j)_z}{v_j-u_i}
+\displaystyle\frac{3 a_i b_j (v_j)_z}{(v_j-u_i)^2}\right\rbrace=0,\quad i=1,\dots,m,\\[6mm]
(b_j)_y&+\displaystyle\sum\limits_{i=1}^m\left\lbrace \left(\displaystyle\frac{a_i b_j}{(v_j-u_i)^2}\right)_x+\left(\displaystyle\frac{a_i b_j (v_j-2 u_i)}{(v_j-u_i)^2}\right)_z\right.\\[6mm]
&\left.-\displaystyle\frac{3 a_i (b_j)_z}{v_j-u_i}+\displaystyle\frac{3 a_i b_j (v_j)_z}{(v_j-u_i)^2}\right\rbrace=0,\quad j=1,\dots,n, 
\end{array}
\end{equation}
that can be brought into Cauchy--Kowalevski form e.g.\ by passing from $t$ to $T=y+t$ with all other variables intact and 
then solving w.r.t.\ $T$-derivatives for all dependent variables. $\Box$

Now let us return to the study of general properties 
of contact Lax pairs (\ref{linearlax-0}). 

First of all, note that 
contact Lax pairs (\ref{linearlax-0}) belong to a broader class of nonisospectral\footnote{Roughly speaking, nonisospectrality here refers to the fact that the Lax pairs in question involve the derivatives with respect to $p$; for nonisospectral Lax pairs in general see e.g.\ \cite{bzm,Dun,aslmp, z} and references therein.}
Lax pairs
\begin{equation}\label{genni}
\begin{array}{l}
\chi_y=K_1 (p,\bu) \chi_x+K_2 (p,\bu) \chi_z + K_3 (p,\bu)\chi_p,\\[2mm]
\chi_t=L_1 (p,\bu) \chi_x+L_2 (p,\bu) \chi_z + L_3 (p,\bu)\chi_p,
\end{array}
\end{equation}
and thus 
at least in principle are amenable to an appropriate version of
the inverse scattering transform, see e.g.\ \cite{aslmp} and references therein, which paves the way to constructing explicit exact solutions of the nonlinear systems admitting such Lax pairs. For examples of general Lax pairs (\ref{genni}) and discussion thereof see e.g. \cite{ck, dfk, krm, mas} and references therein;  for some examples of dispersive (3+1)-dimensional integrable systems see e.g.\  
\cite{F06,sz}.
\looseness=-1

\begin{prop}[\cite{aslmp}]
\label{pro} A system (\ref{sys-gen}) admits a 
contact Lax pair of the form (\ref{linearlax-0}) if and only if it admits a nonlinear Lax pair for $\psi=\psi(x,y,z,t)$ of the form
\begin{equation}\label{nlp-gen}
\psi_y=\psi_z f(\psi_x/\psi_z,\bu),\quad \psi_t=\psi_z g(\psi_x/\psi_z,\bu)
\end{equation}
with the same functions $f$ and $g$ as in (\ref{linearlax-0}). 
\end{prop}

System (\ref{nlp-gen}) is nothing but a pair of nonstationary Hamil\-ton--Jacobi equations of a special form, see e.g.\ \cite{k, aslmp} and references therein for nonlinear Lax pairs of this kind.
Systems like (\ref{nlp-gen}) are a special case of multitime Hamil\-ton--Jacobi systems that were intensively studied in a different context, see e.g.\ \cite{Lio}.\looseness=-1


Let us also point out that using a formal expansion of $\chi$ in $p$ enables one, at least in principle, to find an infinite hierarchy of nonlocal conservation laws for the system under study using (\ref{linearlax-0}), cf.\ \cite{aslmp} for details. Now we proceed to illustrate the above general results by several new examples of (3+1)-dimensional integrable systems with linear contact Lax pairs.\looseness=-1

\begin{example}\label{4ddkp-ex}\normalfont
Let $\bu=(u,v,w,r,s)^\mathrm{T}$,
$f=w p^2+v p+\alpha u$, and $g=w^2 p^3+k v w p^2+s p+r$, where $k$ and $\alpha$ are arbitrary constants. For $k=2$, $\alpha=1$ and $w=1$ the system in question reduces to the
case  of $m=1$, $n=2$ in the first of two classes from Theorem~\ref{conth}, if we identify $u_0\equiv u, u_1\equiv v, v_0\equiv r, v_1\equiv s$; this case is studied in detail in \cite{aslmp}.

The associated nonlinear Lax pair (\ref{nlp-gen}) for the above $f$ and $g$ reads
\[
\psi_y=\psi_z(w(\psi_x/\psi_z)^2+ v \psi_x/\psi_z+\alpha u),\quad \psi_t=\psi_z(w^2(\psi_x/\psi_z)^3+k v w (\psi_x/\psi_z)^2+s\psi_x/\psi_z+r)
\] 
while the linear contact Lax pair (\ref{linearlax-0}) 
has the form 
\be\label{lax-dga} 
\begin{array}{rcl}
\chi_y&=&(2 w p + v) \chi_x + (-w p^2+\alpha u)\chi_z+ (w_z p^3+(v_z-w_x) p^2+(\alpha u_z-v_x) p-\alpha u_x)\chi_p,\\[2mm] 
\chi_t&=&(3 w^2 p^2 +2k v w p+s)\chi_x+
(-2 w^2 p^3-k v w p^2+r)\chi_z\\[2mm] &&+
(2 w w_z p^4+(k v w_z+k w v_z-2 w w_x) p^3+(-k v w_x-k w v_x+s_z) p^2+(r_z-s_x) p-r_x)\chi_p, 
\end{array}
\ee
and the associated integrable system for $\bu$, resulting from the compatibility condition for (\ref{lax-dga}), i.e., from (\ref{zcr}) with the above $f$ and $g$,  reads
\begin{equation}\label{4ddga}
\begin{array}{l}
w^2 (w_x-(k-2) v_z)=0, \\[1mm] 
\alpha u_t-r_y-\alpha r u_z+\alpha u r_z+v r_x-\alpha s u_x=0,\\[1mm] 
v_t-s_y-2 \alpha k v w u_x-r v_z+2 w r_x-s v_x+ v s_x+\alpha u s_z=0,\\[1mm] 
w (w (2 k-3) v_x- s_z+2 \alpha w u_z+k v v_z+2 v w_x-2 w_y+2 \alpha u w_z)=0,\\[1mm] 
w_t-w r_z+2 w s_x-3 \alpha w^2 u_x+\alpha k v w u_z-k v w v_x-k w v_y \\[1mm]
+\alpha k u w v_z+(k v^2-s) w_x-k v w_y+(\alpha k u v-r) w_z=0
\end{array}
\end{equation}

Let us show that this system is an integrable (3+1)-dimensional generalization for (2+1)-dimensional dispersionless KP, dispersionless modified KP, and dispersionless Gardner equations.

First of all, (\ref{4ddga}) admits an integrable reduction $w=1$, and in Example 2 of \cite{aslmp} it is, up to a slight difference in notation, shown that the system obtained from (\ref{4ddga}) upon 
setting $k=2$ in addition to $w=1$ is an integrable (3+1)-dimensional generalization of the dispersionless KP equation. 

Now let us 
show that (\ref{4ddga}) for $k=3/2$ is a (3+1)-dimensional integrable generalization of the 
(2+1)-dimensional dispersionless Gardner and dispersionless modified KP equations, 
see e.g.\ \cite{FMN} and references thertein for those two. 

Upon putting $\boldsymbol{u}_z=0$ in (\ref{4ddga}) with $k=3/2$ we get
\begin{equation}\label{4ddga-3d}
\begin{array}{l}
w^2 w_x=0, \\[1mm] 
\alpha u_t-r_y+v r_x-\alpha s u_x=0,\\[1mm] 
v_t-s_y-3 \alpha v w u_x+2 w r_x-s v_x+ v s_x=0,\\[1mm] 
w (2 v w_x-2 w_y)=0,\\[1mm] 
w_t+2 w s_x-3 \alpha w^2 u_x- (3/2) v w v_x-(3/2) w v_y +((3/2) v^2-s) w_x-(3/2) v w_y=0
\end{array}
\end{equation}
The above system admits a reduction $w=1$ and $v=\beta u$, where $\beta$ is  an arbitrary constant, and then we get 
\begin{equation}\label{4ddga-red}
\begin{array}{l}
\alpha u_t-r_y+\beta u r_x-\alpha s u_x=0,\\[1mm] 
\beta u_t-s_y-3\alpha \beta u u_x+2 r_x+\beta (us_x-s u_x)=0,
\\[1mm] 
2s_x-3 \alpha u_x-(3\beta /2)(\beta u u_x+u_y)=0
\end{array}
\end{equation}
Putting $w=1$, $k=3/2$, and $v=\beta u$ in the above Lax functions $f$ and $g$ we readily observe that
(\ref{4ddga-red}) admits a nonlinear Lax pair (where now $\psi$ depends only on $x,y,t$) of the form
\[
\psi_y=\psi_x^2+\beta u \psi_x+\alpha u, \psi_t=\psi_x^3+(3/2)\beta u \psi_x^2+s \psi_x+r,
\] 
which is nothing but the dispersionless Lax pair for the (2+1)-dimensional dispersionless Gardner equation, see e.g.\ \cite{FMN} for the latter, up to a suitable rescaling of dependent and independent variables and minor differences in notation.

Note that (\ref{4ddga-red}) 
reduces to the modified dispersionless KP equation upon putting $\alpha=0$. 

Thus, (\ref{4ddga}) is an integrable (3+1)-dimensional generalization for the following (2+1)-dimensional integrable systems: dispersionless KP, dispersionless modified KP, and dispersionless Gardner equations.\looseness=-1
\end{example}

\begin{example}\label{4ddsk-exa} \normalfont 
Let $\bu=(u,v,w,r,s,a,b)^\mathrm{T}$, $f=p^3+v p^2+u p+w$ and $g=p^5+2 v p^4+r p^3+s p^2+a p+b$ 
that is 
up to an obvious change of notation the case of $m=3$ and $n=5$ in class I of Theorem~\ref{conth}. 

In analogy with the above, 
the compatibility condition (\ref{zcr}) for the associated contact Lax pair then yields 
the following nonlinear system:
\begin{equation}\label{4ddsk}
\begin{array}{l}
w_t-b_y-a w_x-b w_z+u b_x+w b_z=0,\\[1mm] u_t-a_y -a u_x+u a_x+w a_z-b u_z+2 v b_x-2 s w_x=0,\\[1mm] 3 a_x-2 b_z-r_y-v a_z-3 r u_x+2 r w_z+u r_x+w r_z+s u_z-2 s v_x+2 v s_x -8 v w_x=0,\\[1mm] v_t-s_y+3 b_x-a v_x+2 v a_x-b v_z-v b_z -3 r w_x-2 s u_x+s w_z+u s_x  +w s_z=0,\\[1mm] -2 r_z+4 u_z+v_x+4 v v_z=0,\\[1mm] 2 r v_z-v r_z +6 v u_z -4 v v_x+3 r_x-2 s_z-5 u_x+4 w_z=0,\\[1mm] 2 r u_z-3 r v_x+2 v r_x +s v_z-v s_z +2 u v_x-8 v u_x +6 v w_z+2 w v_z -2 a_z+3 s_x-2 v_y-5 w_x=0
\end{array}
\end{equation}
Upon putting $\boldsymbol{u}_z=0$ in (\ref{4ddsk}) 
we can further impose the reduction 
$v=0$, $w=0$, $s=0$, $b=0$, $r=5u/3$, after which we end up with the 
system 
\begin{equation}\label{4ddsk-3dre}
\begin{array}{l}
u_t-a_y-a u_x+u a_x =0,\\[1mm] 9 a_x-5 u_y-10 u u_x=0,\\[1mm]
\end{array}
\end{equation}
which is nothing but a version of the dispersionless BKP equation, cf.\ for example \cite{FMN} for the latter, which in turn
also is the dispersionless limit of two (2+1)-dimensional integrable equations, namely (2+1)-dimensional versions of Caudrey--Dodd--Gibbon--Sawada--Kotera and Kaup-Kupershmidt equations, cf.\ for instance \cite{ka}, 
Thus (\ref{4ddsk}) is 
a 
(3+1)-dimensional integrable generalization of the dispersionless BKP equation. 
\end{example}

\begin{example}\label{4ddnvn-exa} \normalfont  
Letting 
$\boldsymbol{u}=(u,v,w,r,s)^\mathrm{T}$ and putting
\begin{equation}\label{lax-nvn}
f=u/p+v,\quad g=-(1/3)u^3/p^{3}+r/p^{2}+u w/p+s, 
\end{equation}
yields, 
upon equating to zero the coefficients at all powers of $p$ in  (\ref{zcr}), the system
\begin{equation}\label{4dnvn}
\begin{array}{l}
u_t+u v w_z-2 u w v_z +v w u_z -s u_z+2 u s_z-u w_y-w u_y=0,\\[1mm]  v_t-s_y-s v_z+v s_z =0,\\[1mm]  r_y-2 u^2 w_z-u w v_x+3 r v_z- v r_z +u s_x=0,\\[1mm] 
2 r u_x-u r_x -(2/3) u^3 u_z-u^3 v_x =0,\\[1mm]  2 u r_z +u^2 u_y-u^2 v u_z -3 r u_z+2 r v_x+(4/3) u^3 v_z- u^2 w_x=0,
\end{array}
\end{equation}
which is a 
novel integrable (3+1)-dimensional generalization of (2+1)-dimensional dispersionless asymmetric Nizhnik--Veselov--Novikov equation, see Subcase 2.1 in Subsection 4.1 of \cite{FMN} for the latter and cf.\ also \cite{P} and references therein; 
the said (2+1)-dimensional dispersionless system is recovered when we drop the $z$-dependence 
by putting $\boldsymbol{u}_z=0$ (and $\chi_z=0$)
and 
then  imposing 
the reduction 
$v=s=r=0$.\looseness=-1
\end{example}

\begin{example}\label{4dnlbex}\normalfont Let $\boldsymbol{u}=(v,w,a_1,\dots,a_n,u_1,\dots,u_n)^\mathrm{T}$, when $n$ is any natural number, and
\begin{equation}\label{lax-4dnlb}
f=-p^2/2-v p-w,\quad g=\sum\limits_{j=1}^{n}\displaystyle\frac{a_j}{(p-u_j)}
\end{equation}
Then the compatibility condition (\ref{zcr}) for the associated contact Lax pair yields 
the nonlinear system
\begin{equation}\label{4dnlb}\hspace*{-5mm}
\begin{array}{l}
(a_i)_t =(u_i^2/2 -w) (a_i)_z+3a_i u_i v_z+u_i a_i (u_i)_z -(u_i+v) (a_i)_x-a_i ((u_i)_x+v_x-2w_z),\quad i=1,\dots,n,\\[1mm]   
(u_i)_t = (u_i^2/2 -w) (u_i)_z+u_i^2 v_z
-(u_i+v) (u_i)_x+u_i (w_z-v_x)-w_x,\quad i=1,\dots,n
\\[2mm]   
v_y =\displaystyle\frac12\sum\limits_{i=1}^n (a_i)_z,\\[5mm]   
w_y =\displaystyle \frac12\sum\limits_{i=1}^n \left(u_i(a_i)_z +a_i (u_i)_z\right) 
+2v_z\sum\limits_{i=1}^n a_i -\sum\limits_{i=1}^n (a_i)_x
\end{array}
\end{equation}
which is a  
novel integrable (3+1)-dimensional generalization of generalized Benney system from \cite{z}; 
we recover the latter upon putting $\boldsymbol{u}_z=0$ (and $\chi_z=0$)
and subsequently imposing the reduction $v=0$.\looseness=-1
\end{example}

In 
connection with the above examples let us mention that an integrable 
(3+1)-dimensional generalization of the 
(2+1)-dimensional dispersionless Davey--Stewartson equation was recently found, using the contact Lax pair approach from \cite{aslmp}, 
in \cite{APC}; see also \cite{bls} for some other examples with contact Lax pairs.\looseness=-1

We stress that there also exist integrable (3+1)-dimensional systems with finitely many dependent variables that admit contact Lax pairs whose Lax functions are not rational. In particular, below 
we present a system \cite{asaml} that, 
to the best of our knowledge, is the first example of an integrable (3+1)-dimensional system with a nonisospectral Lax pair 
whose Lax operators are algebraic, rather than 
rational, in the spectral parameter $p$.\looseness=-1


\begin{thm}[\cite{asaml}]\label{pr1}
The seven-component (3+1)-dimensional evolutionary system 
\begingroup
\allowdisplaybreaks
\begin{align}
a_t&=\displaystyle\frac{1}{r^2-2 r s a + 2 s^2 b}\biggl((4 w (r a - s b) - v r) a_x + r a_y
+ \bigl( 2 w \left(2 a(r a - s b) - r b\right) - u r\bigr) a_z\notag\\
&+ (v s - 2 w r) b_x - s b_y + (2 w (s b -r a) + u s) b_z
+(r-s a) u_x + (r a - 2 s b) u_z + (2 s b-r a) v_x\notag\\
&+ 2(a (s b- ra) + r b) v_z + 2(a (a r- s b)-rb) w_x
+ 2\left(2 a^2 (a r - s b) - 3 b a r + 2 s b^2\right) w_z\biggr),\notag\\ 
b_t&=\displaystyle\frac{1}{r^2-2 r s a + 2 s^2 b}\biggl(2 (2 w r-v s) b a_x + 2 s b a_y
+ 2 (2 w (r a -  s b) - u s) b a_z\notag\\ 
& + (2 s(v a - 2 w b) - v r) b_x + (r-2 s a) b_y
+ (2 (u s a - w r b) - u r) b_z\notag\\ 
&+(2 s (b-a^2) + r a) u_x + 2 (r-s a) b u_z
- 2 (r-s a) b v_x - 2 (r a - 2 s b) b v_z\notag\\ 
& + 2 (r a - 2 s b) b w_x + 4 (a (r a - s b) - r b) b w_z
\biggr),\notag\\ 
r_t&=\displaystyle\frac{1}{r^2-2 r s a + 2 s^2 b}\biggl((v s - 2 w r) r a_x - r s a_y
- (2 w (r a - s b) - u s) r a_z+ (2 w r-v s) s b_x\notag\\ 
& + s^2 b_y + (w r^2-u s^2) b_z+(s a - r) s u_x + (2 s b-r a) s u_z
+ (r-s a) r v_x + (r a - 2 s b) r v_z\notag\\ 
&+ (2 s b-r a) r w_x - 2 ( a(r a - s b) - r b) r w_z\biggr),\notag\\[2mm]
s_t&=w_x + a w_z + w a_z,\notag\\
u_t&=a r_x + 2 b r_z - s b_x,\notag\\
v_t&=r_x + a r_z + a s_x + 2 b s_z - s a_x + s b_z,\notag\\
w_t&=s_x + a s_z + s a_z.\notag 
\end{align}
\endgroup
is integrable: 
%
it 
admits a
nonisospectral
Lax pair of the form (\ref{linearlax-0}) with
algebraic 
Lax functions $f$ and $g$ given by
\begin{equation}\label{fgp}
\hspace*{-10mm}
f=u+v p+w p^2+(r+s p) \sqrt{p^2+2 a p+2 b},\quad 
g=\sqrt{p^2+2 a p+2 b},
\end{equation}
namely, 
\[
\hspace*{-2mm}
\begin{array}{rcl}
\chi_y&=&\displaystyle\frac{1}{g}\biggl(\bigl(2 s p^2+(r+3 s a+2 w g)p+r a+v g+2 s b\bigr)\chi_x
+\bigl(-s p^3-(w g+s a) p^2+p r a+2 r b+u g\bigr)\chi_z\\[5mm]
&&+\bigl(s_z p^4+(2 a s_z+s a_z+r_z+g w_z-s_x)p^3
+((v_z-w_x)g+2 b s_z+r a_z+s b_z
-2 a s_x-s a_x\\[2mm]&&-r_x+2 a r_z)p^2 
+((u_z-v_x)g+r b_z-r a_x-s b_x-2 b s_x-2 r_x a+2 b r_z)p
-r b_x-g u_x-2 b r_x\bigr)\chi_p\biggr),\\[5mm]
\chi_t&=&
\displaystyle\frac{1}{g}\biggl((p+a)\chi_x+(a p+ 2b)\chi_z+(a_z p^2+p (b_z-a_x)-b_x)\chi_p\biggr),
\end{array}
\]
and a nonlinear Lax pair of the form (\ref{nlp-gen}) with $f$ and $g$ given by (\ref{fgp}):
\[ 
\hspace*{-3mm}
\begin{array}{rcl}
\psi_y&=&u\psi_z +v\psi_x+w\psi_x^2/\psi_z
+(r\psi_z+s\psi_x) \sqrt{(\psi_x/\psi_z)^2+2 a\psi_x/\psi_z+2 b},\\[3mm]
\psi_t&=&\psi_z\sqrt{(\psi_x/\psi_z)^2+2 a\psi_x/\psi_z+2 b}. 
\end{array}
\]
\end{thm}

\section{Outlook}

In the present article we have reviewed a recent construction \cite{aslmp} for a large class of integrable (3+1)-dimensional systems with Lax pairs involving contact vector fields, thus showing inter alia that integrable (3+1)-dimensional systems are by no means as scarce as it was believed for a long time. Let us stress that the class in question contains inter alia infinitely many integrable (3+1)-dimensional systems 
admitting Lax pairs of the form (\ref{linearlax-0}) with the Lax functions $f$ 
and $g$ rational in $p$, as well as the first known example \cite{asaml} of a (3+1)-dimensional integrable system with a nonisospectral Lax pair whose Lax operators are
algebraic, rather than rational, in the variable spectral parameter $p$. Moreover, using the construction in question 
we have found in Examples \ref{4ddkp-ex}--\ref{4dnlbex} above four novel integrable (3+1)-dimensional systems which generalize the following 
(2+1)-dimensional integrable systems: dispersionless modified KP, dispersionless Gardner, dispersionless BKP, and dispersionless asymmetric Nizhnik--Veselov--Novikov equations, and the
generalized Benney system.\looseness=-1

In closing let us point out 
two open problems related to the above results 

\begin{itemize}
    \item Can one find an example of integrable system with contact Lax pair (\ref{linearlax-0}) such that the associated Lax functions $f$ and $g$
    are transcendental, rather than algebraic or rational, in $p$? 

\item Is it possible to find examples of integrable (3+1)-dimensional systems with contact Lax pairs (\ref{linearlax-0}) such that the systems in question can be brought into Cauchy--Kowalevski form and the vector $\bu$ of unknown functions has just two or three components? 
\end{itemize}

{\protect\vspace*{-5mm}}
\subsection*{Acknowledgments} The author would like to thank the anonymous referees for useful suggestions and the Ministry of Education, Youth and Sports of the Czech Republic (M\v{S}MT \v{C}R) for kindly providing support. Some computations in the present paper were performed with the aid of the package Jets \cite{BM} for Maple\textsuperscript{\textregistered},
whose use is hereby gratefully acknowledged. 
{\protect\vspace*{-3mm}}
\subsection*{Declarations}
\noindent{\bf Funding}  Open access publishing supported by the institutions participating in the CzechELib Transformative
Agreement. This research was supported in part by 
the Ministry of Education, Youth and Sports of the Czech Republic (M\v{S}MT \v{C}R) through RVO funding for I\v{C}47813059. 

\smallskip

\noindent{\bf Data availability} Data availability statement is not applicable to the present paper.

\smallskip

\noindent{\bf Conflicts of interest} The author is unaware of any conflicts of interest.
%
{\protect\vspace*{-5mm}}
%


\end{document}